\documentclass[twocolumn,showpacs,preprintnumbers,amsmath,amssymb]{revtex4}
\usepackage[dvips]{graphicx,color}
\usepackage{dcolumn}
\usepackage{bm}
 
\usepackage{ulem}

\begin{document}
\def\bea{\begin{eqnarray}}
\def\eea{\end{eqnarray}}
\def\be{\begin{equation}}
\def\ee{\end{equation}}
\def\rra{\right\rangle}
\def\lla{\left\langle}
\def\rv{\bm{r}}
\def\la{\Lambda}
\def\sgm{\Sigma^-}
\def\eps{\epsilon}
\def\ms{M_\odot}
\def\bc{B=100\;\rm MeV\!/fm^3}
\def\beff{B_{\rm eff}(\rho_B)}
\def\sc{\sigma_{\rm crit} \approx 70\;\rm MeV\!/fm^2} 
\def\qc{\rho_{\rm ch}}
\def\fv{f_V}

\preprint{APT/123-QED}

\title{Hot hadron-quark mixed phase including hyperons}%

\author{Nobutoshi Yasutake$^1$}
 \email{yasutake@th.nao.ac.jp}
\author{Toshiki Maruyama$^2$}
\author{Toshitaka Tatsumi$^3$}
\affiliation{%
$^1$Division of Theoretical Astronomy, National Astronomical Observatory of Japan, 2-21-1 Osawa, Mitaka, Tokyo 181-8588, Japan\\
$^2$Advanced Science Research Center, Japan Atomic Energy Agency, Tokai, Ibaraki 319-1195, Japan\\
$^3$Department of Physics, Kyoto University, Kyoto 606-8502, Japan
}%

\date{\today}

\begin{abstract}
We study the hadron-quark phase transition with the finite size effects at finite temperature.
For the hadron phase, we adopt  a realistic equation of state in the framework of the Brueckner-Hartree-Fock theory including hyperons. 
The properties of the mixed phase are clarified by considering the finite size effects under the Gibbs conditions. 
We find that the equation of state becomes softer than that at zero-temperature for some density region. 
We also find that the equation of state gets closer to that given by the Maxwell construction. 
Moreover, the number of hyperons is suppressed by the presence of quarks.
These are characteristic features of the hadron-quark mixed phase, and should be important for many astrophysical phenomena such as mergers of binary neutron stars.
\end{abstract}

\maketitle
\section{\label{sec:level1}Introduction}

Nowadays the effects of quark matter on various astrophysical phenomena have been studied extensively. For example, the cooling of compact stars have been studied in Ref.~\cite{page00, blaschke00, blaschke01, grigorian05}. Other examples include the effects of quark matter on gravitational wave radiation~\cite{lin06, yas07, abdikamalov08}, neutrino emissions~\cite{hatsuda87, nakazato08, sagert08}, rotational frequencies~\cite{burgio03}, and the energy release during the collapse from neutron stars to quark stars~\cite{yas05, zdunik07}, etc..
In particular, the mechanisms of supernovae and gamma-ray bursts  have not been clearly understood; the QCD phase transition may take place in such phenomena and take critical role~\cite{gentile93}.

However, there are left many uncertainties for the hadron-quark phase transition, e.g. the equation of state (EOS) of quark matter or deconfinement mechanism. Assuming the quark deconfinement transition to be of first order, we find it causes a thermodynamical instability and the mixed phase appears around the critical density. 
Since there are two conserved quantities, baryon number and electric charge, the phase equilibrium in the mixed phase must be 
carefully treated by applying the Gibbs conditions \cite{glendenning92}, instead of the Maxwell construction.
A simple treatment of the mixed phase may be the bulk Gibbs calculation, where phase equilibrium of two bulk matter is considered without electromagnetic interaction and surface tension.
Generally the properties of the mixed phase should strongly  depend on electromagnetic interaction and surface tension, and these  effects, sometimes called ``{\it the finite-size effects}", lead to the non-uniform "Pasta" structures. 
The EOS of the mixed phase becomes similar to the one under the bulk Gibbs calculation for weak surface tension, and to the one given by the Maxwell construction for strong surface tension~\cite{voskresensky03,endo06,maruyama07}. The charge screening is also important for their mechanical instability. 

In the previous papers these finite-size effects have been properly taken into account to elucidate the properties of the pasta structure and demonstrate the importance of the charge screening at zero temperature \cite{maruyama07}. However, finite temperature comes in many cases such as relativistic heavy-ion collisions and astrophysical phenomena. 
In this paper, we study the hadron-quark mixed phase with the finite-size effects at finite temperature by extending the previous works of Maruyama et al.~\cite{maruyama07}.
We adopt the Brueckner-Hartree-Fock EOS by Baldo et al. for the hadron phase~\cite{baldo98}. 
The EOS includes hyperons as well as nucleons, interacting with hadronic two-body forces and nucleonic three-body forces. 
For the quark phase, we adopt the thermodynamic bag model for simplicity.
We impose the Gibbs conditions on the phase equilibrium, taking into account the finite-size effects. 

This paper is organized as follows. In Sec.~II, we outline our framework. In Sec.~III, we present numerical results. Sec.~IV is devoted to the conclusion and discussion where we give some astrophysical implications of our results.
\section{Equation of state}
\subsection{\label{sec:level2} Equation of state for hadron phase at finite temperature 
---Brueckner-Hartree-Fock theory}
Our theoretical framework for the hadron phase of matter
is the nonrelativistic Brueckner-Hartree-Fock approach~\cite{bhf} 
based on the microscopic nucleon-nucleon~($NN$), nucleon-hyperon~($NY$), and hyperon-hyperon~($YY$) 
potentials. 
The Brueckner-Hartree-Fock calculation is a reliable and well-controlled theoretical approach 
for the study of dense baryonic matter.
Detailed procedure can be found in Refs.~\cite{baldo98,hypmat,vi00}.

We adopt the Argonne $V_{18}$ potential~\cite{v18} for $NN$ potentials, 
and semi-phenomenological Urbana UIX nucleonic three body forces~\cite{uix}     
and the Nijmegen soft-core NSC89 $NY$ potentials \cite{nsc89}.
Unfortunately, there are not reliable $YY$ potentials now, because no $YY$ scattering data exist. 
Therefore, we neglect $YY$ potentials in this paper.
Recently, the maximum mass of neutron stars are calculated \cite{schulze06} 
using the NSC97 $NY$ and $YY$ potentials \cite{nsc97}. 
However, their maximum masses are quite similar.

In this paper, we discuss the hadron-quark mixed phase with finite-size effects at finite temperature
\footnote{In realistic situations neutrinos may be also trapped inside matter, but we hereafter consider the neutrino-free case, leaving the neutrino-trapped case in a future paper.}.
At finite temperature, there is no microscopic calculation.
To take into account the effects of finite temperature, we adopt \textit{Frozen Correlations Approximation}~\cite{baldo99, nicotra06a,nicotra06b}.
In this approximation, the correlations at finite temperature are assumed to be the same with the ones at zero temperature. 
It is found to be good accuracy at finite temperature
by past studies~\cite{baldo99, nicotra06a,nicotra06b}. 
Accordingly, we focus on hadron-quark mixed phase at the finite temperature 
and adopt this approximation  for the neutrino-free case.

In the following, we briefly describe our framework.
First we get the chemical potential $\mu_i$  from the number density $n_i$, 
\begin{eqnarray}
n_i     &=& \cfrac{g}{(2\pi)^3} \int^\infty_0 f_i(p) ~ 4 \pi p^2 dp     \\
f_i(p) &=& \cfrac{1}{ {\rm exp} \{ (\varepsilon_i-\mu_i)/ T \} +1},
\end{eqnarray}
where  $\varepsilon_i$ and $f_i(p)$ are the single-particle energy and the Fermi-Dirac distribution function, respectively, whereas subscript $i$ shows the particle species, $i=n,p,\Lambda,\Sigma^-$. 
We set each degeneracy factor $g=2$, and adopt each mass as $m_n=m_p=939$ MeV, $m_\Lambda=1115.7$ MeV, and $m_{\Sigma^-}=1197.4$ MeV.
We note that $\varepsilon_i$ includes the interaction energy $U_i$ as well as the kinetic energy \cite{baldo99, baldo04}, 
\begin{eqnarray}
\varepsilon_i = \sqrt{m_i^2 + p^2 } + U_i.
\end{eqnarray}

Finally, we get the free-energy density $\mathcal{F}$ as following   
\begin{eqnarray}
\mathcal{F}_H = &&\sum_i  \left\{ \cfrac{g}{(2\pi)^3} \int^\infty_0 \sqrt{m_i^2+p^2} f_i(p) ~4 \pi p^2 dp + \cfrac{1}{2} ~ U_i n_i \right\} \notag \\
&&-Ts_H ,
\label{eq:free}
\end{eqnarray}
where $s_H$ is the entropy density calculated from
\begin{eqnarray}
s_H = &&- \sum_i \cfrac{g}{(2\pi)^3} \int^\infty_0 \{ f_i(p) {\rm ln}f_i(p) +(1-f_i(p))  \notag \\
&& \times {\rm ln}(1-f_i(p)) \} ~ 4 \pi p^2 dp.
\end{eqnarray}

The total pressure for uniform hadron phase is given by
\begin{eqnarray}
P_H= \sum_i \mu_i n_i   -  \mathcal{F}_H. 
\end{eqnarray}

\subsection{\label{sec:mit} Equation of state for quark phase \\
---The thermodynamic bag model}

Unfortunately, the current theoretical description of quark matter 
includes many uncertainties, seriously limiting the predictability of EOS at high baryon density. 
For the time being, we only resort
to a phenomenological model for the quark matter EOS 
and try to put the constraint of parameters 
by a few experimental information. 

In this paper, we adopt the thermodynamic bag model to construct EOS for simplicity. 
The bag model may be an oversimplified model for quark matter.  It should be interesting to compare our results with those given by other models; 
e.g. Nambu-Jona-Lasinio model~\cite{burgio08, yasutake09} or Polyakov-Nambu-Jona-Lasinio model~\cite{fukushima08a, fukushima08b}.  

We get the number density $n_Q$, the pressure $P_Q$, and the energy density $\eps_Q$,
\begin{eqnarray}
n_Q&=&\sum_q \cfrac{g}{(2\pi)^3} \int^\infty_0 f_q(p) 4 \pi p^2 dp, \\ 
\epsilon_Q&=&\sum_q \cfrac{g}{(2\pi)^3} \int^\infty_0 \sqrt{m_q^2 + p^2 } f_q(p) 4 \pi p^2 dp +B, \\ 
s_Q &=& - \sum_q \cfrac{g}{(2\pi)^3} \int^\infty_0 \{ f_q(p) {\rm ln}f_q(p) +(1-f_q(p)) \notag \\
 && \times {\rm ln}(1-f_q(p)) \} ~ 4 \pi p^2 dp \\
\mathcal{F}_Q&=& \epsilon_Q -Ts_Q \\
P_Q&=& \sum_q \mu_q n_q   -  \mathcal{F}_Q. 
\end{eqnarray}
for uniform quark matter,
where $f_q(p)$ is the Fermi-Dirac distribution function of the quark $q$ ($=u,d,s$), $m_q$ is its current mass, 
and $B$ is the energy density difference between 
the perturbative vacuum and the true vacuum, the bag constant.
In this paper, we employ $B=100$~MeV fm$^{-3}$, 
and the degeneracy factor $g = 6$.
We use massless $u$ and $d$ quarks and $m_s = 150$ MeV,
and ignore any anti-particles
because quark chemical potential are much larger than masses and temperature. 

The baryon number is conserved as
\bea
 n_B &=& \frac{1}{3}\left(n_u + n_d + n_s\right) =  \frac{1}{3}n_Q \:.
\label{e:baryon} 
\eea

\subsection{Hadron-quark mixed phase under the Gibbs conditions}
\label{s:mix}

\begin{figure}[htb]
\includegraphics[width=0.48\textwidth]{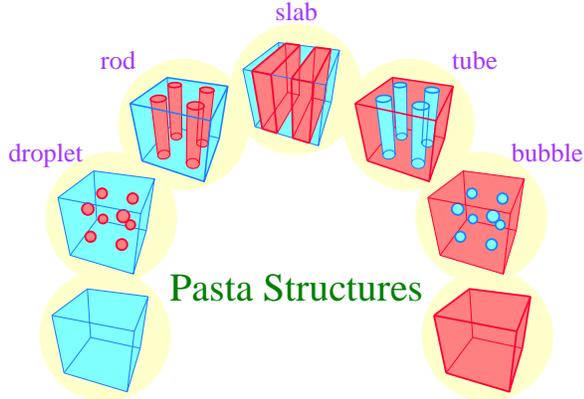}
\caption{Pasta structures in the Wigner-Seitz approximation. We assume that one of them appears in the hadron-quark mixed phase under the periodic condition.}
\label{fig:pasta}
\end{figure}

To take into account the finite-size effects, 
we impose the Gibbs conditions on the mixed phase~\cite{heiselberg93},
which require the pressure balance and the equality of the chemical potentials
between two phases besides the thermal equilibrium.
We employ the Wigner-Seitz approximation in which the whole space 
is divided into equivalent cells with given geometrical symmetry, 
specified by the dimensionality 
$d=3$ (droplet or bubble), $d=2$ (rod or tube), or $d=1$ (slab).
The structures of tube and bubble are opposite distributions of rod and droplet as illustrated in Fig.~\ref{fig:pasta}. 

The quark and hadron phases are separated in each cell with volume $V_W$ :  
a lump made of the quark phase with volume $V_Q$ is embedded
in the hadronic phase with volume $V_H$ or vice versa.
A sharp boundary is assumed between the two phases and the surface energy
is taken into account in terms of a surface-tension parameter $\sigma$.
The surface tension of the hadron-quark interface is poorly known, 
but some theoretical estimates based on the MIT bag model 
for strangelets \cite{jaf} and
lattice gauge simulations at finite temperature \cite{latt} suggest
a range of $\sigma \approx 10$--$100\;\rm MeV\!/fm^2$.
We show results using $\sigma=40\;\rm MeV\!/fm^2$ 
in this article.

We use the Thomas-Fermi approximation for the density profiles of
hadrons and quarks.
The Helmholtz free energy for each cell is then given as
\be
F = \int_{V_H}\!\!\! d^3\rv\, \mathcal{F}_H[n_i(\rv)] 
 + \int_{V_Q}\!\!\! d^3\rv\, \mathcal{F}_Q[n_q(\rv)] + F_e + E_C + E_S
\ee
with $i=n,p,\la,\sgm$, $q=u,d,s$, 
$\mathcal{F}_H$~($\mathcal{F}_Q$) is the Helmholtz free energy density for hadron~(quark) matter, 
and $E_S=\sigma S$ the surface energy with $S$  being the hadron-quark interface area. 
$F_e$ is the free energy of the electron gas.
For simplicity, muons are not included in this paper.
The value of $E_C$ is the Coulomb interaction energy calculated by,
\be
 E_C = \frac{e^2}{2} \int_{V_W}\!\!\! d^3\rv d^3\rv'\,
 \frac{   n_{\rm ch}(\rv)  n_{\rm ch}(\rv')   }{    |\rv-\rv'|     }    \:,
\ee
where the charge density is given by 
\be
e n_{{\rm ch}} (\rv) = \sum_{i=n,p,\la,\sgm,e} Q_i n_i(\rv)
\ee
in $V_H$ and 
\be
e n_{{\rm ch}} (\rv) = \sum_{q=u,d,s,e} Q_q n_q(\rv)
\ee
in $V_Q$ with $Q_i$ (or $Q_q$) being the particle charge 
($Q=-e < 0$ for the electron).
Accordingly, the Coulomb potential $\phi(\rv)$ is defined as
\be
 \phi(\rv) = -\int_{V_W}\!\!\! d^3\rv'\, 
 \frac{   e^2 n_{\rm ch}(\rv')   }{   \left| \rv - \rv' \right|   } + \phi_0 \:,
\label{e:vcoul}
\ee
where $\phi_0$ is an arbitrary constant representing 
the gauge degree of freedom.
We fix it by stipulating the condition,  
$\phi(R_W) = 0$, 
as before \cite{maruyama05,voskresensky02,tatsumi03}.
The Poisson equation then reads
\bea
 \Delta \phi (\rv) = 4 \pi e^2 n_{{\rm ch}}(\rv) \:.
\label{e:poisson}
\eea

Under the Gibbs conditions, we must consider chemical equilibrium at the hadron-quark
interface as well as inside each phase;
\bea
 && \mu_u+\mu_e = \mu_d = \mu_s \:, 
\nonumber\\
 && \mu_p+\mu_e = \mu_n = \mu_\la = \mu_u + 2\mu_d \:, 
\nonumber\\
 && \mu_{\sgm} + \mu_p = 2\mu_n \:.
\label{e:chemeq}
\eea
For a given baryon number density 
\be
 n_B = \frac{1}{V_W}\! \left[
 \sum_{i=n,p,\la,\sgm} \int_{V_H}\!\!\!\! d^3\rv n_i(\rv)
 +\! \sum_{q=u,d,s} \int_{V_Q}\!\!\!\! d^3\rv \frac{n_q(\rv)}{3} \right] \:,
\ee
Eqs.~(\ref{e:poisson}--\ref{e:chemeq}), 
together with the global charge neutrality condition,
$\int_{V_W}\!\!{d^3\rv} n_{{\rm ch}}(\rv)=0$, 
obviously fulfill the requirement by the Gibbs conditions.

\section{Numerical Results}
\subsection{Thermal effects on EOS}

\begin{figure*}[htb]
\includegraphics[width=0.48\textwidth]{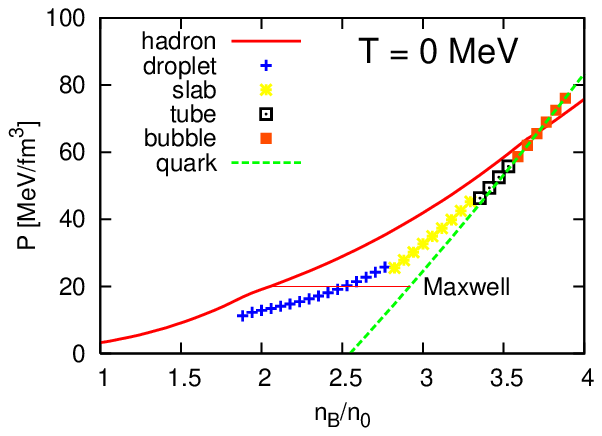}
\includegraphics[width=0.48\textwidth]{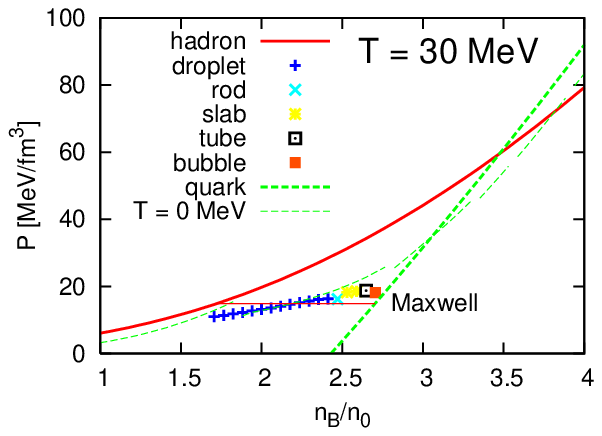}
\\
\includegraphics[width=0.48\textwidth]{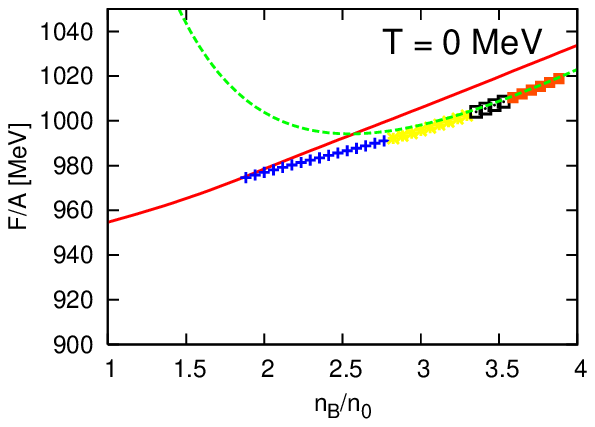}
\includegraphics[width=0.48\textwidth]{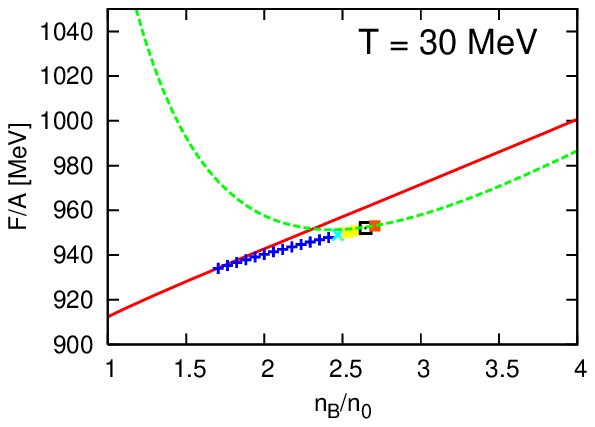}
\caption{
(Color online)
EOS of the mixed phase~(thick dots) in comparison with pure hadron and quark phases~(thin curves).
The upper panels show the pressure at zero temperature~(left), and $T=30$~MeV~(right).
We also show, for comparison, the mixed phase 
by the Maxwell construction by thin solid line.
Lower panels: 
The free energy per baryon $F/A$ of the mixed phase (thick curves)
in comparison with pure hadron and quark phases (thin curves).
The left panels show the zero temperature case, and 
the right panels the finite temperature case~($T=30$~MeV).
Different colored segments of the mixed phase are chosen by minimizing the energy.
}
\label{fig:eos}
\end{figure*}

Using above relations, we study the hadron-quark mixed phase at finite temperature.
Upper panels of Fig.~\ref{fig:eos} indicate  
the resulting pressures of the hadron-quark mixed phase with that of the pure hadron and quark phases
over the relevant range of baryon density at zero temperature~(the upper left panel) 
and finite temperature~(the upper right panel). 
Here, we take $T=30$~MeV. 
The thin curves indicate the pure hadron and quark phases,
while the thick dots indicate the mixed phase 
in its various geometric realizations by the full calculation.
The transitions between the different geometrical structures are, by construction, discontinuous 
and a more sophisticated approach would be required for a more 
realistic description of this feature.
We also show, for comparison, the hadron-quark phase transition 
by the Maxwell construction by the thin solid line.

\begin{figure*}[tb]
\includegraphics[width=0.48\textwidth]{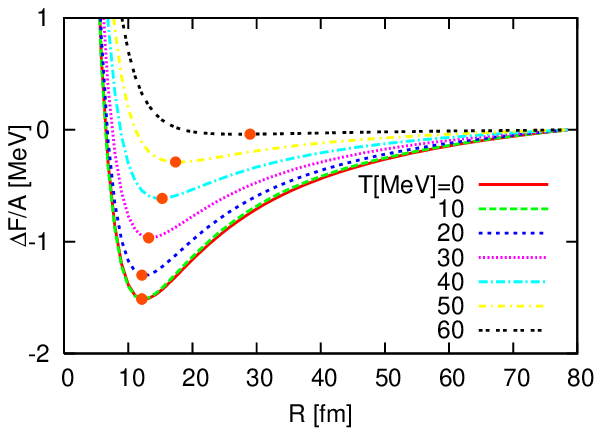}
\includegraphics[width=0.48\textwidth]{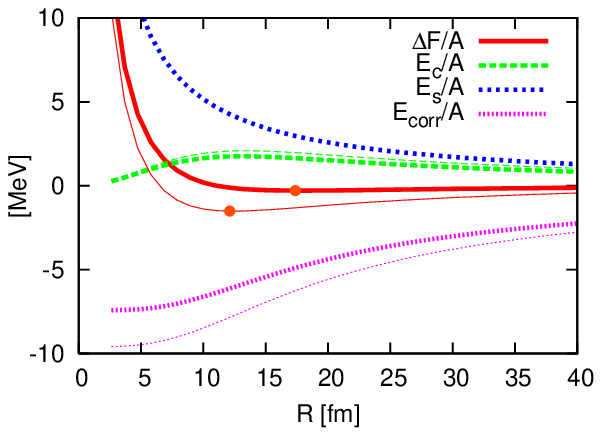} 
\caption{
(Color online) 
The Left panel shows that the droplet radius $R$ dependence of the free energy per baryon for $n_B=2 n_0$ and different temperatures. 
The quark volume fraction $(R/R_W)^3$ is fixed to be the optimal value at $T=0$~MeV for each curve. 
The free energy is normalized by the value at $R\rightarrow\infty$.
Filled circles on each curve shows the energy minimum. The results are for $B=100$ MeV/fm$^3$, $\sigma=40$ MeV/fm$^2$. 
The Right panel shows each contribution to the $R$ dependence of the free energy, the Coulomb energy, the surface energy or the correlation energy per baryon~($E_{\rm c}/A$, $E_{\rm s}/A$, $E_{\rm corr}/A$) at $T=0$ MeV(thick lines) and $T=50$ MeV(thin lines). For the definite meaning of the correlation energy, see the text. 
}
\label{fig:stable}
\end{figure*}

Compared with the zero temperature case, the mixed phase is restricted and 
EOS gets close to that by the Maxwell construction,
 though we properly apply the Gibbs conditions. 
The restriction of the mixed phase has been already demonstrated at $T=0$ due to the charge screening effect \cite{maruyama07}. 
We can see that the geometrical structure also becomes unstable due to the thermal effects. 
The left panel of Fig.~\ref{fig:stable} shows the free energies per baryon 
of the droplet structure 
at several values of temperature. The quark volume fraction $(R/R_W)^3$ is fixed to exclude the trivial $R-$ dependence. Here we use, for example, the optimal value of $(R/R_W)^3$ at $T=0$~MeV in every curve. 
We normalize them by subtracting the free energy at infinite 
radius,  
 $
  \Delta F = F(R)-F(R\rightarrow\infty),
 $
 to show the $R$ dependence clearly. 
The structure of the mixed phase is mechanically stable below $T \sim 60$ MeV, 
but the optimal value of the radius $R$ is shifted to the larger value as $T$ increases.
This behavior is a signal of the mechanical instability and comes from the charge screening effect and the thermal effect.

To elucidate this point more clearly, we show the each contribution to $\Delta F/A$  in the right panel of Fig.~\ref{fig:stable}; 
i.e.\ the Coulomb energy per baron $E_{\rm C}/A$, the surface energy per baryon $E_{\rm S}/A$, 
and the correlation energy per baryon $E_{{\rm corr}}/A$.
The baryon density and the temperature are set as  $n_B=2 n_0$ and $T=50$ MeV. 

When we treat the Coulomb potential and charge densities in a self-consistent manner, we can see the charge screening effect; 
it gives rise to the Debye screening mass for the Coulomb interaction and induces the rearrangement of charge densities (see Fig.~4). 
Then $E_{\rm C}/A$ 
exponentially decreases as $R$ increases for large $R$. 
At finite temperature the value of $E_{\rm C}$ 
decreases due to the hyperon mixing in comparison with the zero temperature case, but it is a little effect. 
On the other hand thermal effect mainly emerges through the rearrangement of charge densities; 
the kinetic and strong-interaction energies give rise to a new $R$ dependence, 
which is called the correlation energy in Ref.\cite{voskresensky02}, 
proportional to the difference of charge configuration from the uniform one in the absence of 
the Coulomb interaction. 
It has been shown that $E_{{\rm corr}}/A$ exhibits $-R^{-1}$ behavior for large $R$ \cite{voskresensky02}. 
Its magnitude decreases as $T$ increases due to the reduction of the Coulomb interaction.
Thus the optimal value of $R$ is shifted to a larger value at finite temperature. 
Note that the extreme case such that $R(R_W)\rightarrow \infty$ corresponds to the Maxwell construction for bulk matter. 
In our formulation, the pasta structures disappear at $T \gtrsim 60$ MeV.

The pressure of the mixed phase at finite temperature is lower than that at zero temperature as shown in the upper right panel of Fig.~\ref{fig:eos}, 
while the pressure of the pure hadron or quark phase at finite temperature becomes higher than that at zero temperature. 
When we write the pressure $P$ as the sum of the cold part~(zero temperature) $P_{\rm cold}$ and the thermal part $P_{\rm thermal}$,  
$
P=P_{\rm cold}+P_{\rm thermal},
$
usually we find $P_{\rm thermal}>0$. On the other hand, we find $P_{\rm thermal}<0$ for the mixed phase. 
Similar results have been shown by previous studies,
 though they did not include hyperons nor the finite-size effects\cite{burgio08,yasutake09}.
This behavior is characteristic for the mixed phase.
How does such characteristic behavior of pressure appear?

\begin{figure*}[t]
\includegraphics[width=0.48\textwidth]{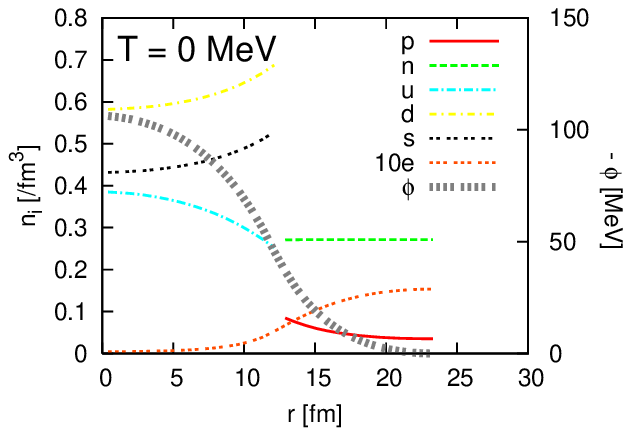}
\includegraphics[width=0.48\textwidth]{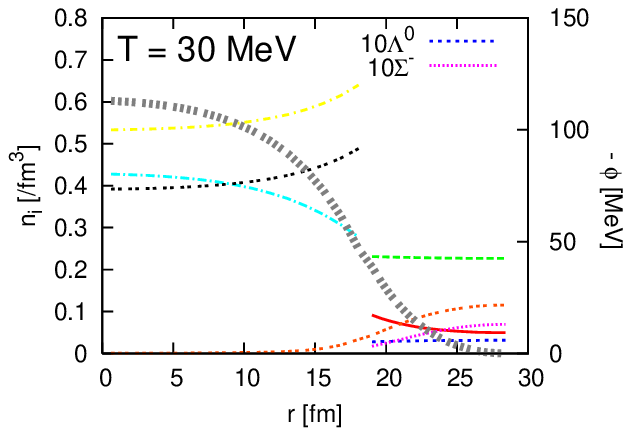}
\caption{
(Color online)
Density profiles and Coulomb potential $\phi$ within 3D~(quark droplet) for $n_B=2 n_0$ at zero temperature~(left panel) and finite temperature, $T=30$~MeV~(right panel). 
The cell sizes are $R_W= 23.7$~fm~($T=0$~MeV) and $R_W= 28.8$~fm~($T=30$~MeV).
The droplet radii are $R= 12.5$~fm~($T=0$~MeV) and $R= 18.7$~fm~($T=30$~MeV).
}
\label{fig:prof}
\end{figure*}

Since the degrees of freedom for quark matter is larger than those for hadron matter, 
the change of heat in quark matter becomes larger than that in hadron matter as temperature is increased. 
Hence the free energy change of quark matter is larger than that of hadron
matter (compare the lower panels of Fig.~\ref{fig:eos}).
This means that the quark phase is favored rather than the hadron phase in the mixed phase for a given density
 at finite temperature.  
Hence, the mixed phase contains more quarks at finite temperature.
The free energy is decreased by the inclusion of more quarks and this change exceeds the increase of the kinetic energy.
As a result, EOS of the mixed phase becomes softer at finite temperature.
We can see this behavior in Fig.~\ref{fig:prof}. 
It shows the density profiles within 3D cell (quark droplet) for $n_B=2 n_0$ at zero temperature~(left panel) and finite temperature, $T=30$~MeV~(right panel).
Clearly, the volume fraction of quark matter at finite temperature is larger than that at zero temperature.  

Another point is that the onset 
density~of the mixed phase is shifted to a lower value (compare the left panels of Fig.~\ref{fig:eos} with the right panels). 
This behavior also comes from the fact that  the quark phase is favored rather than the hadron phase for a given density at finite temperature, as discussed above. 

\subsection{Hyperon suppression in the mixed phase}

\begin{figure*}[t]
\includegraphics[width=0.48\textwidth]{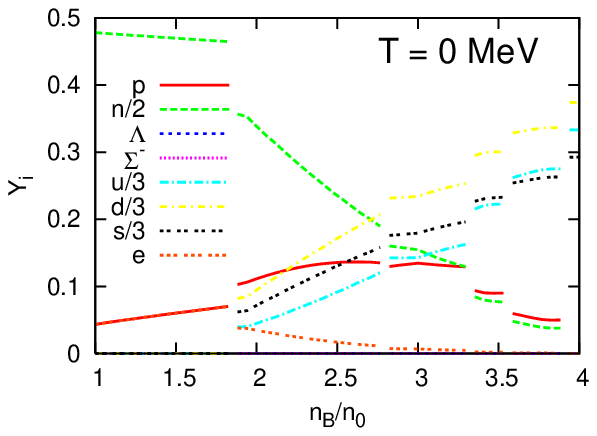}
\includegraphics[width=0.48\textwidth]{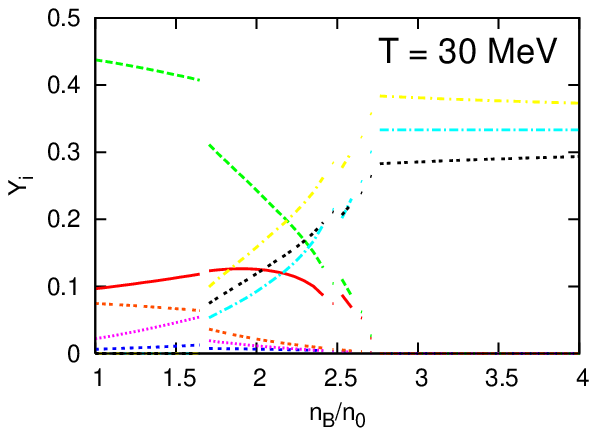} \\
\includegraphics[width=0.48\textwidth]{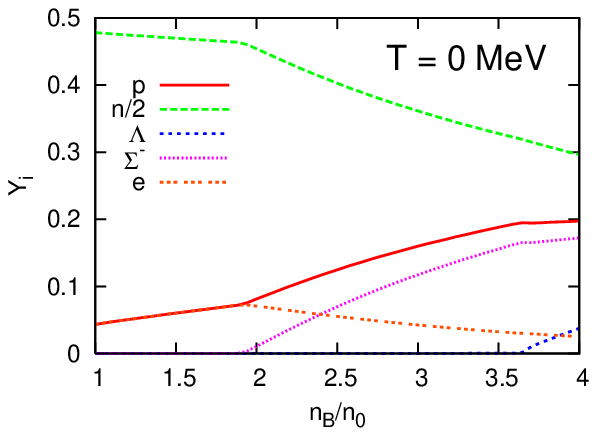}
\includegraphics[width=0.48\textwidth]{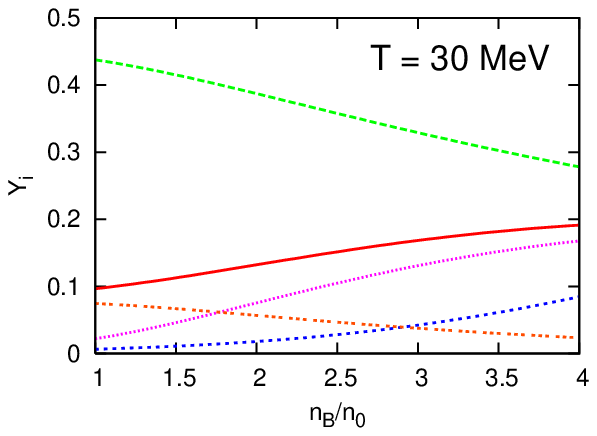} \\
\caption{
(Color online)
Particle fractions at zero temperature~(left panels), and the finite temperature, $T=30$ MeV~(right panels).
The upper two panels include quark phase and the mixed phase as well as hadron phase. 
Lower panels: Particle fractions in the hypothetical pure hadron matter.}
\label{fig:Yi}
\end{figure*}

Particle fractions of quark and hadron species are shown in Fig.~\ref{fig:Yi};
upper panels show them during the phase transition, while the lower panels for hypothetical pure hadron matter for comparison.
In the previous papers hyperon suppression has been demonstrated in the mixed phase at zero temperature~\cite{maruyama07} 
(see upper and lower left panels), 
since the properties of hadron matter inside the mixed phase is much different from those of pure hadron matter. 
Actually hadron matter is positively charged and its density becomes lower in the mixed phase, both of which disfavor the mixture of hyperons. 
At finite temperature hyperon mixture becomes rather easy since the Fermi seas of hadrons are diffused by the thermal effect and entropy is 
increased \cite{prakash97} (see lower right panel); the hyperons appear at low density~($\sim n_0$) at $T=30$ MeV, 
while they do 
above $2n_0$ at $T=0$ (the lower left panel). 
However, we can see that hyperons are also suppressed in the mixed phase even in such a case 
for 
the same reason as at zero temperature (see upper right panel).
\footnote{There is a controversy about the $\Sigma^-$-n interaction. 
The recent experimental result about hypernuclei has suggested that it is repulsive~\cite{noumi02,saha04}, 
while we used a weak but attractive interaction in this paper. 
In this case $\Sigma^-$ first appears. 
It would be interesting to see how our results are changed in this case, and we will discuss it in the future work.}

\subsection{Maximum-mass problem}

\begin{figure*}[t]
\includegraphics[width=0.96\textwidth]{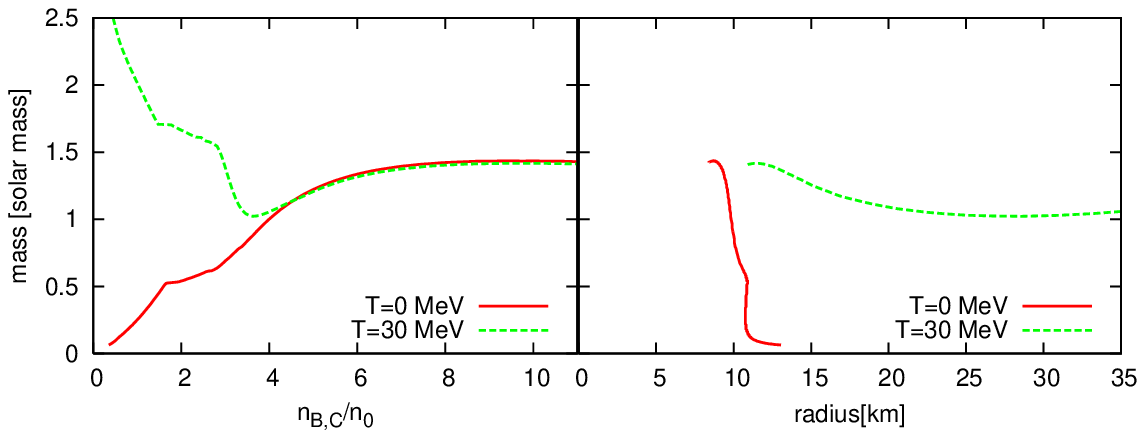}
\caption{
(Color online)
Mass-central density~(left panel) and mass-radius relations~(right panel).
The solid lines show them at zero temperature, and the dashed lines  at $T=30$ MeV .
}
\label{fig:mr}
\end{figure*}

At the last of this section, we briefly discuss some implications of our results on the maximum-mass problem. It is well-known that EOS becomes too soft to support the canonical mass of $\sim 1.4M_\odot$, once hyperons are taken into account above the nuclear density~\cite{baldo06}. 
The deconfinement transition to quark matter has been suggested to circumvent the difficulty~\cite{schulze06, baldo06}. 
We have confirmed this suggestion in the previous paper \cite{maruyama07} 
by treating the hadron-quark mixed phase in a proper way. 
This argument has been done for $T=0$ case, but similar discussion is possible within the present framework.
We show the mass - central density and mass - radius relations for isothermal hybrid stars in FIG.~\ref{fig:mr},
obtained by solving the Tolman-Oppenheimer-Volkoff equations. 
Below the sub-nuclear density~($n < 0.1$ fm$^{-3}$), we use EOS by the Lattimer \& Swesty~\cite{lattimer91}.
We can see that 
the maximum mass $M_{\rm max}$ at finite temperature is slightly smaller than that at zero temperature; 
$M_{\rm max}=1.42 M_{\odot}$ at $T=30$ MeV and  $M_{max}=1.43 M_{\odot}$ at zero temperature. 
This result is easily understood by considering that hybrid stars in the maximum-mass region 
are predominantly governed by the EOS at high-density, where pure quark matter is realized; 
the ratio of temperature to chemical potential at several times the nuclear density is 
much less than 1 for $T<30$ MeV, which makes the EOS a little stiff.
Note again that matter considered in this paper is neutrino-free, 
while neutrino-trapped and/or isentropic case would be more interesting for, e.g., supernovae. 
However, these figures may give us some suggestions about the stability of newly born neutron stars after the mergers of neutron star binaries. 

\section{Conclusions and Discussions}
We have studied the hadron-quark phase transition at finite temperature. 
We have taken into account the {\it finite-size effects}  imposing the Gibbs conditions on the phase equilibrium, and calculating the density profiles in a self-consistent manner. 

At finite temperature, EOS of the hadron-quark phase transition gets close to that given by the Maxwell construction.
It is due to the mechanical instability of the geometrical structure induced by the thermal effect.  
Pressure of the mixed phase at finite temperature are 10--30 \% smaller than that at zero-temperature 
though the similar behavior appears without hyperons~\cite{burgio08, yasutake09}. 
This behavior is characteristic of the hadron-quark mixed phase 
which can be explained by
the large degree of freedom in the quark phase.
Hyperon fractions are suppressed by the appearance of the mixed phase, as in the case of zero temperature.

Our calculations are subject to the  neutrino-free~(low $Y_l$) case at finite temperature. Such situation will appear in mergers of neutron star-neutron star binaries or black hole-neutron star binaries~\cite{shibata06}, and our result may change their dynamical aspects. Of course, we should take into account isentropic and $Y_l\neq 0$ situation in the core of supernovae. This work is now in progress.

Finally we note again that EOS has many uncertainties, especially for quark matter. 
We simply adopted the thermodynamic bag model in this paper, and 
used the density-temperature independent bag constant and surface tension, while it would be interesting to include such dependence for a realistic description~\cite{baldo03, voskresensky09}.
Moreover, chiral restoration or color super conductivities may also change our results~\cite{kashiwa07,kashiwa07b,jorge07,fukushima08,yasutake09}.
These are open questions for astrophysics and nuclear physics.

\begin{acknowledgments}
We are grateful to S.~Chiba, H.~J.~Schulze for their warm hospitality and fruitful discussions.
This work was partially supported by the 
Grant-in-Aid for the Global COE Program 
``The Next Generation of Physics, Spun from Universality and Emergence''
from the Ministry of Education, Culture, Sports, Science and Technology
(MEXT) of Japan and the Grant-in-Aid for Scientific Research (C) (20540267, 21105512, 19540313).
\end{acknowledgments}

\newpage 
\bibliography{yas09c}

\end{document}